\documentclass[]{mn2e}

\usepackage{amssymb}
\usepackage{graphicx}
\usepackage{amsmath}
\usepackage{graphics}

\usepackage{color}

\title[Visible continuum and bands in the ISEC]{On the visible continuum and bands in the interstellar extinction curve}

\author[R. Papoular]{Renaud Papoular$^{1}$\thanks{E-mail:papoular@wanadoo.fr}\\
$^{1}$Service d'Astrophysique and Service de Chimie Moleculaire,\\
CEA Saclay, 91191 Gif-s-Yvette, France}

\begin{document}

   \maketitle
\label{firstpage}

\begin{abstract}
This work purports to help understand the InterStellar Extinction Curve in and near the visible range. In this range, crystalline materials are known to be transparent, so amorphous dust is needed. Molecular modeling experiments are used to compute the electronic spectra of various, relatively large, carbon and silicate structures. Hardly any transition shows up beyond 0.4 $\mu$m when the structure is in its ground state (the lowest, most stable state, usually crystalline). This is no longer the case as soon as the structure is distorted in any way.  Examples of simulated distortions (or ``defects'') are: angular or linear bond alteration, insertion of free radicals near the main structure, dangling bonds; their cumulative effects lead to the amorphous state. It is shown that, in this state, a structure bears a majority of weak transitions and a minority of strong ones. As the structure grows in size, the former ultimately form a weak continuum already detected experimentally, in the visible, on amorphous carbons and silicates. The stronger transitions will manage to emerge above the continuum, especially when they bunch together by chance near the same wavelength. Parallels are drawn between several properties of the computed continua and transitions and the observed continuum and Diffuse Interstellar Bands.
.

\end{abstract}


\section{Introduction}

The InterStellar Extinction Curve (ISEC) due to dust absorption extends over the whole electromagnetic range, between the near UV and the millimeter domains. The range between the 2175-\AA  bump and the IR vibrations is devoid of spectral features, except for  the seemingly isolated Diffuse Interstellar Bands (DIBs) which emerge weakly above a continuum; the latter declines down to the vibration spectral range with its strong silicate features at 10 and 18 $\mu$m. In this range, a weak polarization bump, the Serkowski peak has been abundantly documented.
The present work is devoted to the understanding of the intensity and shape of that portion of the ISEC. A great deal of published work has been devoted to the DIBs, but no consensus over their origin seems to have been reached yet. By contrast, it is widely believed that the underlying continuum must be carried by amorphous materials. 

A detailed study of \it amorphous hydrogenated silicon \rm (a-Si:H) was provided by Street \cite{str}, and explains the general physical concepts of a semi-conductor and of the disorder which characterizes amorphous states. A similar study of graphite, amorphous carbon and \it  hydrogenated amorphous carbons \rm (a-C:H or HAC) had previously been made by Robertson \cite{rob}. Both are essential for the present purposes, as are the more recent and seminal books by Kuzmany \cite{kuz} and Egerton \cite{ege}.

However, to my knowledge, amorphous silicates have raised less interest , except for the laboratory work of Scott and Duley \cite{sco}. Amorphous carbons were studied with a wealth of experimental techniques, but theoretical publications seem to be wanting. This work purports to study the absorption and bands of such materials from first principles.

 The electronic spectrum of many candidate dust materials, in their crystalline form, presents a $gap$ between their $valence$ and \it conduction bands\rm . In insulators, this gap is large and often referred to as the \it transparency range \rm of crystals. In pure and perfect crystals, light absorption in that range is extremely weak and hardly measurable. If, however, atomic impurities are present, such as iron or chromium ions, they create new electronic states which, in turn, give rise to electronic transitions in the visible range. More generally, any structural defect lowers its symmetry and has similar effects. Defects include coordination defects such as bond deformation and dangling bonds, interstitial inclusions (adatoms or admolecules), vacancies (missing atoms).

As a result of the accumulation of defects in amorphous materials, and of the crowding of accompanying absorption features, a weak, continuous, absorption becomes measurable in the gap. Exceptionally strong features may emerge from this underlying continuum.

\begin{figure*}
\mbox{
\includegraphics[width=9cm]{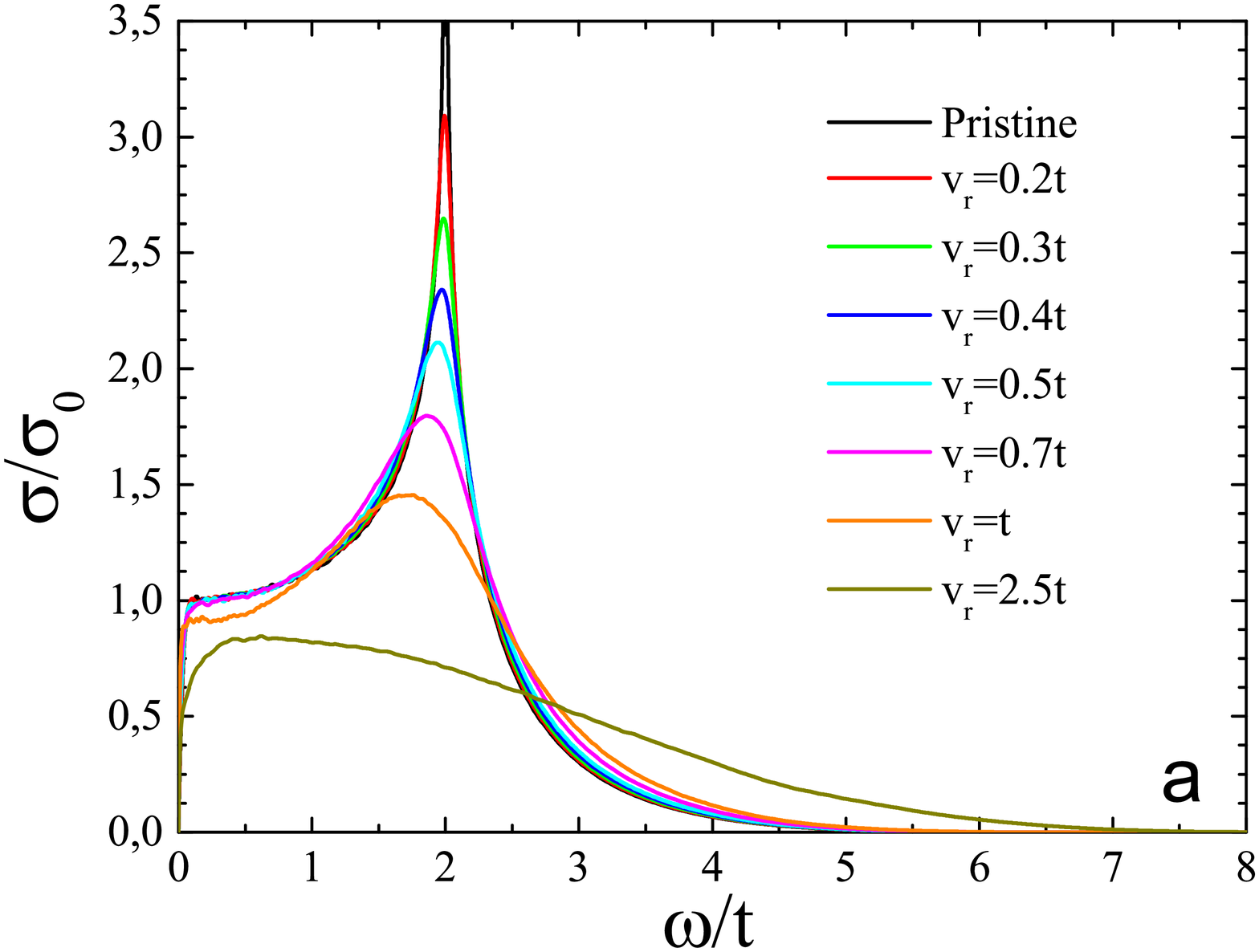}
\includegraphics[width=9cm]{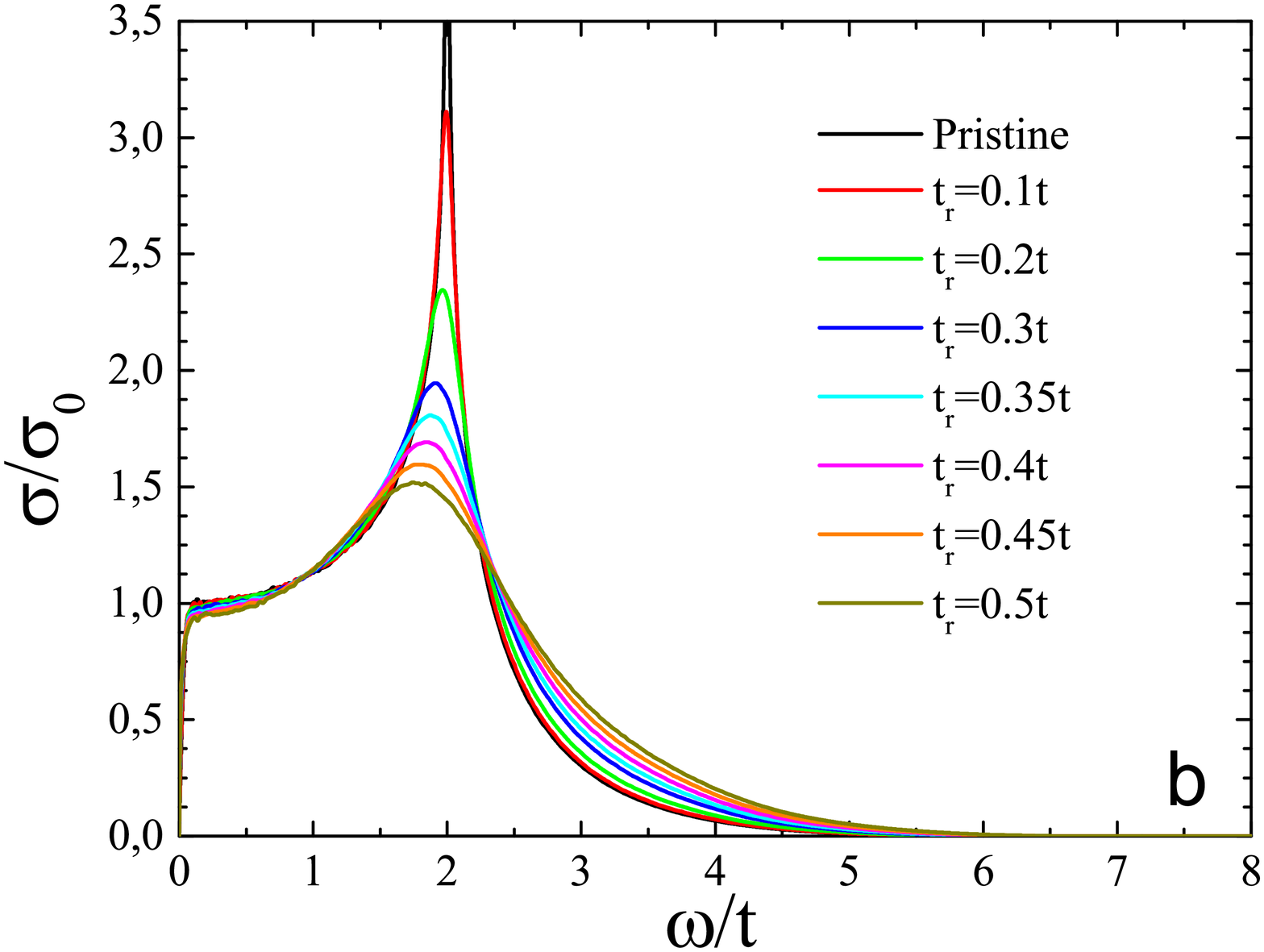}
}
\mbox{
\includegraphics[width=9cm]{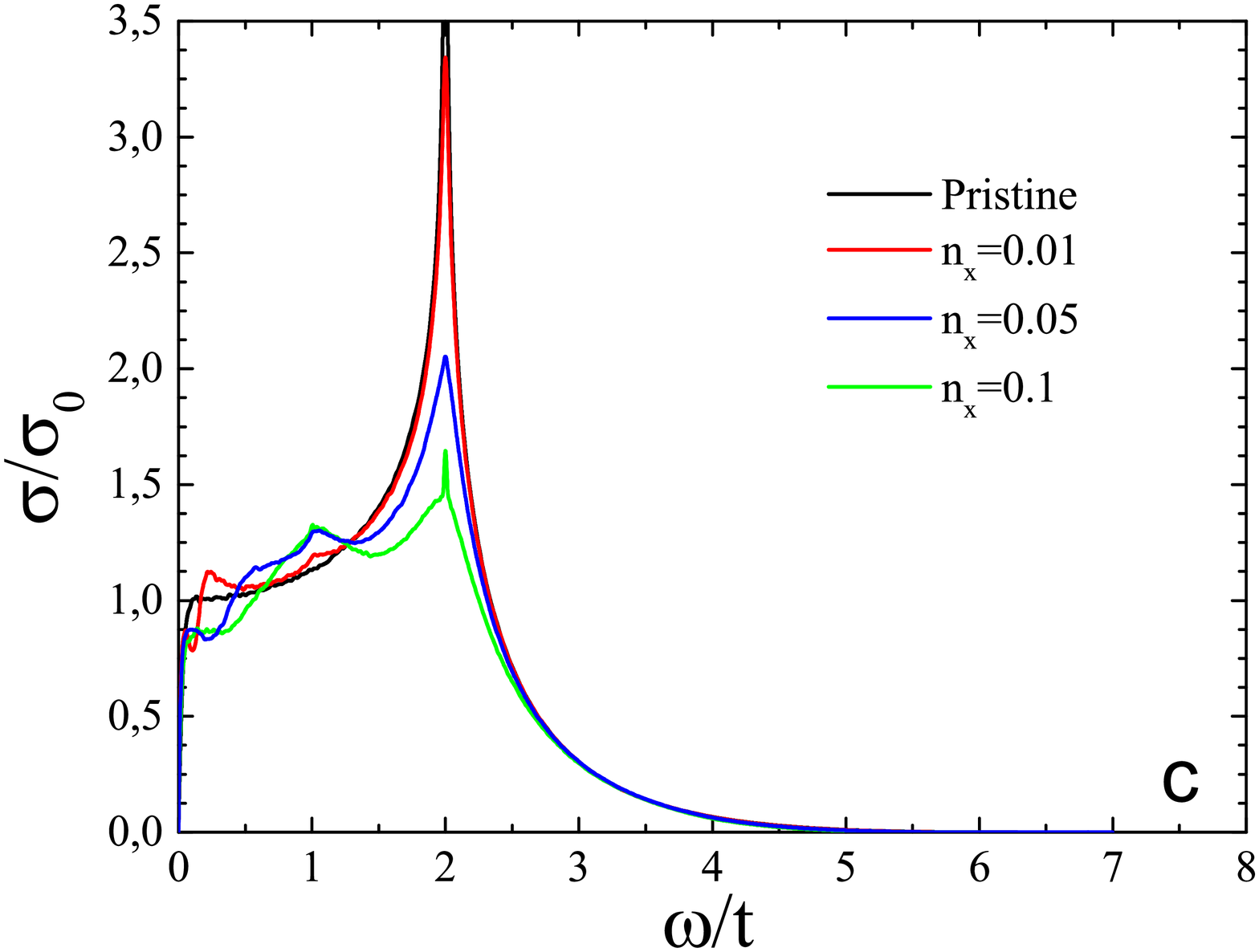}
\includegraphics[width=9cm]{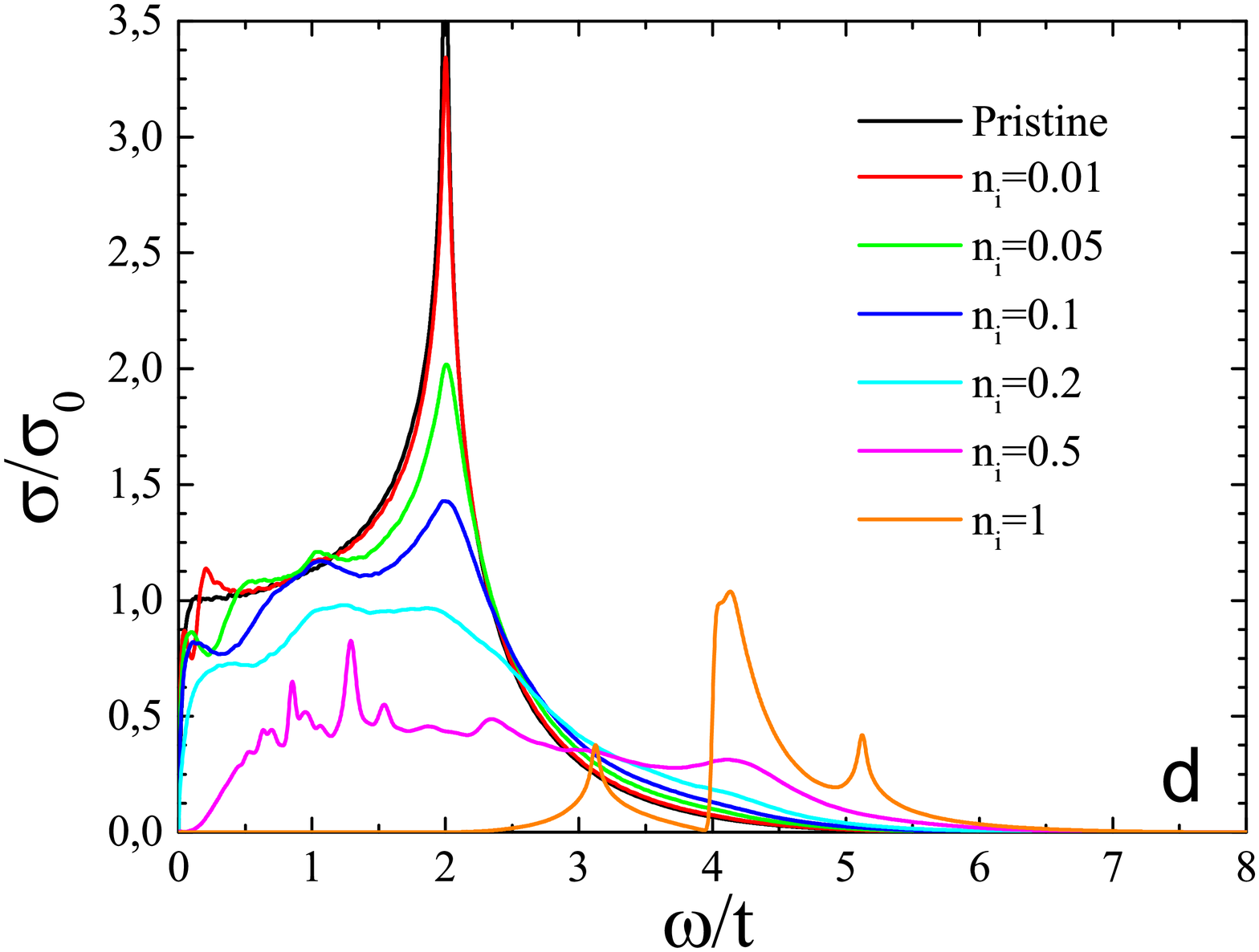}
}
\caption[]{ Optical conductivity $\sigma(\omega)$ of SLG with different kinds of disorder, in units of $\sigma_0=\pi e^2/2h$. (a) Effects of random on-site potentials on $\sigma(\omega)$, mimicking the inhomogeneous distributions of electrical charges. The on-site potential is let to randomly vary in the range $[-v_r,v_r]$, for $v_{r}=0$, 0.2, 0.3, 0.4, 0.5, 0.7, 1, 2.5 (in units of $t$). (b) Effects of random hopping constants on $\sigma(\omega)$, simulating random variations of inter-atomic distances and directions . The hopping integral is let to vary randomly between $[t-t_r,t+t_r]$, for $t_{r}=0$, 0.3, 0.35, 0.4, 0.45, 0.5 (in units of $t$). (c) Effect of random vacancies (missing atoms) in the lattice, with the concentration per carbon atom $n_{x}$ taking the values 0, 0.01, 0.05, 0.1. (d) Effects of random hydrogen-like impurities (adatoms or admolecules) with the concentration per carbon atom $n_{i}$ being 0, 0.01, 0.05, 0.1, 0.2, 0.5, 1.0. For all the plots, the value of the corresponding disorder parameter $v_r$, $t_r$, $n_x$ and $n_i$ increases in order of decreasing sharpness of the peak at $\omega=2t$. Here, $t=2.1$ eV. The visible range of $\omega/t$ extends roughly between 0.5 and 1.5.
}
\label{Fig:OpCond}
\end{figure*}

The formation of dust in space by accretion of atoms (see Gail and Sedlmayr 2013), makes it likely that, whatever its chemical composition, it should be mostly amorphous or highly disordered. A telling experimental demonstration of a disorder effect in candidate dust models was given by Scott and Duley \cite{sco}, who deduced the complex index of refraction of amorphous $forsterite$ (Mg$_{2}$Si$_{2}$O$_{4}$) and \it enstatite \rm  (MgSiO$_{3}$ or Mg$_{2}$Si$_{2}$O$_{6}$) from reflection measurements in the range 0.1 to 20 $\mu$m. They found that, in the transparency range 
($\sim$0.3 to $\sim$3 $\mu$m) of these two non-iron bearing silicates, there remains some weak extinction with different trends in strength as a function of wavelength. This was put to use by Papoular \cite{pap18} to model the so-called Serkowski peak, a wide spectral bump which peaks in the visible and is so weak that it has only been documented in measurements of star light polarization (see Spitzer 1978, Whittet 2003).

The width of the 2175-\AA  feature and its variations are other instances where structural defects were invoked as a possible cause. This feature is usually assigned to graphite grains, but its width is large (about 1 $\mu$m$^{-1}$) and it varies by a factor 2 from source to source in a way that  is not compatible with perfect graphite crystal. An interdisciplinary team investigated whether this behavior could be explained by imperfections in small graphitic structures and found that it is
 compatible with the presence of random vacancies and/or random inclusions (Papoular et al. 2013). A by-product of this work was the evidence for the formation of  narrower absorption features \it in the visible\rm . It is therefore worthwhile to recall here some quantitative results of that work, as illustrated in Fig. \ref{Fig:OpCond}.

This displays the electrical conductance, $\sigma$ of graphene (a layer of graphite), in units of its theoretical DC conductance, $\sigma_{0}=\pi e^{2}/h$, against the energy of the incident photons, whose electric field is parallel to the plane of the layer. We recall that the imaginary part of the associated dielectric function is $\epsilon''=\sigma/\omega$ and the relative efficiency factor is $Q/a\propto \epsilon\omega=\sigma$. Each frame is dedicated to one type of defect (see the caption); each curve corresponds to a different abundance of that defect. The 2175 resonance is displayed, sitting upon a continuum extending over the visible and near UV. This  continuum is much stronger in graphite than in silicates (in the same spectral range) because of the covalent bonds between C atoms, which make it a semi-metal.

All four types of defects reduce the intensity of the 2175 resonance and increase its width; however, the fourth (atomic inclusions) produces particularly strong effects: a distinct lowering of the continuum and the appearance of new narrow features all over the visible spectral range, protruding above the continuum.

Following the course of these considerations, the present work adopts still another approach to the effects of dust defects on the ISEC. Consider a dust particle as a big molecule. In principle, the electron trajectories in any molecular structure can be analyzed into orbitals of increasing energy, using the MO (Molecular Orbital) approximation in its LCAO (Linear Combination of Atomic Orbitals) version. The number of required orbitals increases like the number of constitutive atoms, and with their complexity. Each orbital can harbor at most 2 electrons which must have opposite spins (Pauli's exclusion principle). In equilibrium (ground state), the electrons of the system crowd into the lower ($occupied$) orbitals: the most stable configurations are the most tightly bonded. Higher, $unoccupied$ (virtual) orbitals are needed to describe $excited$ states.

This system of orbitals is the precursor of the valence (or $\pi$) and conduction (or $\pi*$) bands in solids. The higher occupied molecular orbital (HOMO) is separated from the lower unoccupied molecular orbital by a gap, which is usually of order several electron-volts. According to Koopmans Theorem, the absolute value of the HOMO's potential is equal to the ionization potential (IP), which is experimentally true to 1 or 2 eV. The orbital energies extend from about -50 to about 10 eV.
 
An incident $visible$ photon may be absorbed if it can lift an electron from one orbital to a higher one (only single-electron transitions will be considered). This is subject to 3 conditions:
 
 a) The orbital energy difference must be of order 1 eV. For this to be the case, the number of orbitals must be high, i.e. \it the system must be large enough, \rm typically at least several tens of atoms.
 
 b) The transition must satisfy the selection rules: 1) the state spin (sum of orbital spins) S and multiplicity (r= 2S+1) must be conserved; 2) the total orbital quantum number L must change by $\pm1$. 
 
  c) Before excitation, the initial orbital must harbor at least one electron, and the final, at most one. 
  
  A transition is said to be $allowed$ if it satisfies these conditions. The sum of all allowed transition strengths for a single valence electron is 1 (Sum Rule). Other transitions are $forbidden$.
  
   For a given elementary composition of the structure, the energy distribution of the orbitals depends on the atomic configuration (Walsh's rules). In general, in the ground state (electric neutrality, geometric symmetry  and maximum binding energy) of the structure, all the lower orbitals are fully occupied (with electron pairs of opposite spins), so no transition is possible between them. An electron can only migrate from one of these orbitals by crossing the `` gap'' into one of the upper, unoccupied orbitals; transitions are therefore confined to the UV range, and are singlets (S=0, r=1). By contrast, \it in amorphous structures, symmetries are broken and these exclusions are relaxed, to a larger extent the stronger the deformation or the degree of disorder; consequently, transitions appear further into the visible and near IR \rm. Simple geometric distortion only changes the energy distribution of orbitals and the spectral distributions of electronic transitions (transfer of oscillator strength). Atomic inclusions also increase the numbers of orbitals and of possible transitions.

 The vocabulary of defects is borrowed from solid state physics, in which a great deal of effort has gone into the experimental and theoretical study of disordered structures, as recalled above. It can, however be extended to the realm of molecules. In molecular physics, one starts from the ground state and then, for example, distorts the molecule or changes the bond lengths, or creates dangling bonds. The two approaches are complementary, but the molecular approach is perhaps more figurative, effective and flexible.
 
 Of course, a small, isolated structure distorted out of its ground state and left alone will tend by any means towards a lower, more stable state. However, if several such structures cluster together, Van der Waals forces can freeze them into a higher energy and less stable state, a precursor of an amorphous solid or a plastic polymer.
 
Since relevant experimental data on distorted molecules are hardly available, I turned to chemical modeling applied to large molecules having chemical composition similar to those of amorphous hydrogenated carbon (a-C:H or HAC) and silicates (Forsterite and Enstatite). In all cases, increased disorder led to new electronic bands from the near UV to the near IR.

Section 2 below describes the code used here and the computational procedure. This code delivers the properties of the orbitals, the electric dipole moments and the wavelengths and oscillator strengths of transitions. The latter are related to the \it equivalent widths \rm (EW) through the column density of carriers. 

In Sec. 3, the spectrum of a coronene molecule stripped of its H atoms is computed and shown to  lack lines in the visible/near IR. Four versions of this molecule were perturbed differently to simulate disorder and amorphous structures; their computation delivered a large number of new lines, extending the spectrum as far as the near IR. In Sec. 4, the same behavior is demonstrated with Enstatite and Forsterite cells. 

In both cases, orbital energies and spectral distributions of oscillator strengths, absorbance and k index are displayed.

Finally, Sec. 5 summarizes the computed results and draws a parallel between some of these and the corresponding observed properties of IS continuum and DIBs.

\section{Modeling code and computational procedure}

The calculations were made using the Hyperchem package, v8.0, 64 bit, released by Hypercube, Inc., and implemented on a desk-top PC equipped with a Pentium (r)II processor (450 MHz) and MMX (TM) technology. Most state-of-the-art chemical codes are available with this package, from Molecular Mechanics to various \it ab initio \rm methods. A useful comparative guide to the use of these methods was written by Hehre et al. \cite{heh}

Only one of these codes was used in the present work: the semi-empirical PM3 code, developed by J. Stewart \cite{ste} along the same line as AM1. The semi-empirical methods use a rigorous quantum-mechanical formalism occasionally combined with empirical parameters obtained from comparison with experimental results. PM3 incorporates a much larger number and wider variety of experimental data than AM1. These codes compute approximate solutions of Schroedinger's equation, using some form of SCF (Self-Consistent Field) methods, such as the HF (Hartree-Fock) procedure. When the quantum mechanical calculations are too difficult or too lengthy, parameters of the method are taken from experimental data. This sometimes makes them more accurate than poor \it ab initio \rm methods, and they are always faster and can handle larger systems. 

A key feature of the code for our present purposes is the determination of the orbital occupancy of the electrons and calculation of the energies of the electronic orbitals that are necessary to properly describe the total molecular wave function of a given configuration of atoms. The number of orbitals considered in this code is equal to the number of valence electrons in the structure: 1 H for atoms, 2 for He atoms, 4 for all main group atoms and 9 for transition metals. When optimized by minimizing the total energy of the system, the configuration describes the ground state of the structure. In this ``reference configuration'', the orbitals are devided into lower, occupied orbitals and upper,  unoccupied (or ``virtual'') ones. The basic principles underlying this calculation are clearly different from the simple crystal field theory used to describe solid state systems. 

In order to allow a proper calculation of the UV/vis spectrum, one has to go beyond the SCF approximation by taking interactions between single electrons into account. The code uses the CI (Configuration Interaction) procedure for this purpose (see Foresman et al. 1995). Here the term ``configuration'' designates one distribution of available valence electrons over the orbitals. In the CI procedure, ``excited configurations'' are obtained by transfering one or more electron from occupied to unoccupied orbitals. The calculation yields a set of improved molecular states, each of which is represented by a linear combination of these configurations. This procedure is also implemented in the ``Gaussian'' chemical simulation package.

For a given molecule the whole computational procedure begins with the construction of the selected structure on the screen. The constituent atoms are picked up in the periodic table and stuck roughly relative to one another. Adjacent atoms are then linked by single or multiple bonds, as the case may require. As a preparatory step, a Molecular Mechanics modeling code is invoked, which approximately but quickly places the various atoms in a lattice such that the empirical chemical laws governing bonds are obeyed. One then switches to the semi-empirical code to ``optimize'' the structure according to quantum mechanics, meaning that the 3 sets of space coordinates are successively changed so as to decrease the total energy computed at each step, until it reaches a minimum (hopefully the absolute one) to some preset residual error. 

At this stage the code has saved all the information defining the ground state of the structure: geometrical parameters, kinetic energies of nucleons and electrons, repulsion energies between nucleons and between electrons, attraction energies between nucleons and electrons. The CI procedure is then launched. Within the quantum rules governing electrons, and the user-specified maximum incident photon energy (the ``excitation energy''), the code selects the largest possible set of electronic configurations, each being defined by a different way in which an electron in a lower orbital is lifted to a higher one; each configuration generates one band (or transition).

This procedure delivers, in particular, the electron orbital energies and spin occupancies, the electric dipole moment, oscillator strength ($f$) and frequency ($\nu$) for each transition. A transition is possible between two orbitals if the final orbital is not completely occupied. The initial and final orbitals can be identified, as well as the degeneracy of each transition. \it The transition wavelengths are less reliable than their relative distribution in the spectrum. \rm

The code automatically applies the selection rules recalled in the Introduction. Moreover, in crystals, only momentum-conserving, or $direct$, transitions are allowed, because photons do not carry enough momentum for an $indirect$ transition to occur via interaction with a photon. Their oscillator strength is generally higher than $10^{-6}$. Forbidden transitions are orders of magnitude weaker. In the case of amorphous structures, the momentum conservation rule is relaxed (see Jackson et al. 1985). All oscillators are then found to be stronger than, say, $10^{-6}$.

Because the available computer memory is limited, the choice of a value for the excitation energy must be an educated compromise between the size of the molecule and extension of the computed spectral window into the near UV and IR. In the case of carbon structures, one is guided by the fact that the graphite $\pi-\pi*$ transition band extends to photon energies up to about 10 eV, when the sum rule reaches the saturation level for \it one valence electron per C atom \rm. For the $\sigma-\sigma*$ transition, the corresponding numbers are 40 eV and 4 electrons (Taft and Philipp 1965). In order to allow for the computation of structures with 20-30 C atoms, we shall select an excitation energy of 10 eV (unless specified otherwise), since we are mainly interested, here, in the band gaps or visible range. It must be kept in mind, however, that this may artificially restrict the extension of the computed spectrum on both sides.

The finite temperature of the illuminating source must also be taken into account, for it determines the maximum excitation energy. Thus, the effective temperature of O5 stars is 35000 K, which limits the excitation energy to below 10 eV.

The connection of the modeling results with astronomical data is made through the definition of the oscillator strength

\begin{equation}
f=1.2\,10^{12}\frac{\alpha\Delta\nu}{n_{a}}\,,
\end{equation}

where $\alpha$ is the absorbance (absorption per centimeter), $\Delta\nu$ the transition width in cm$^{-1}$ and $n_{a}$ the  density of absorbers in cm$^{-3}$.  This is easily transformed into

\begin{equation}
f=1.2\,10^{12}\frac{EW (\mathrm{cm})}{N_{a}(\mathrm{cm}^{-2})\lambda(\mathrm{cm})^{2}}=1.2\,10^{17}\frac{EW(m\AA{\ })}{N_{a}(\mathrm{cm}^{-2})\lambda(\AA{\ })^2}\,,
\end{equation}

where $EW=\tau\Delta\lambda$ is the equivalent width of the feature, $\tau$, the central optical depth, $\lambda$ the wavelength and $N_{a}$ is the column density of absorbers. By absorber is meant, here, one given computed structure; $n_{a}$ is the inverse of the volume of the absorber.

The next step for our present purposes is to somehow introduce defects in the structure in an attempt to simulate an amorphous structure. This is done here by a) displacing the atoms individually on the screen, b) severing a bond joining two adjacent atoms and c) adding a naked atom (a radical) in the vicinity of the structure. This will be applied first to carbon, then to silicate structures.

\section{Amorphous carbons}
\begin{figure}
\resizebox{\hsize}{!}{\includegraphics{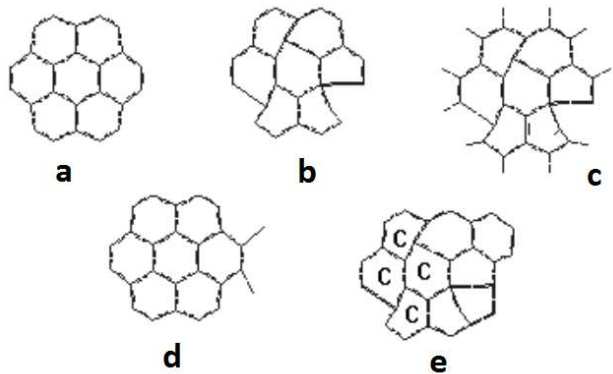}}
\caption[]{Chemical simulations of defects and amorphous structures. The figure 
 displays sketches of 5 of these, which carry the different types of defects considered in Sec. 2: a) a coronene molecule stripped of its peripheral H atoms, but otherwise devoid of defects, b) a distorted coronene, c) a distorted coronene with 12 peripheral H atoms, d) a coronene with 1 single C-C bond and 2 peripheral H atoms and e) a composition of 6- and 5-membered cycles with 4 interstitial C atoms.}
\label{Fig:carbons}
\end{figure}

A large number of structures were studied by means of the code described above. Figure \ref{Fig:carbons} displays sketches of 5 of these, which carry the different types of defects considered in Sec. 2: a) a coronene molecule stripped of its peripheral H atoms, but otherwise devoid of defects, b) a distorted coronene, c) a distorted coronene with 12 peripheral H atoms, d) a coronene with 1 single C-C bond and 2 peripheral H atoms and e) a composition of 6- and 5-membered cycles with 4 interstitial C atoms.

\begin{figure}
\resizebox{\hsize}{!}{\includegraphics{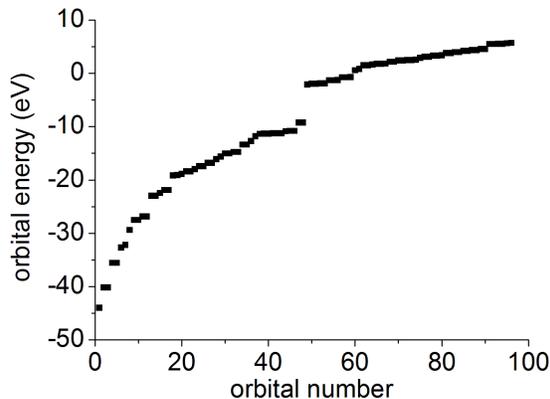}}
\caption[]{Orbital energies of the perfectly ordered structure drawn in Fig. \ref{Fig:carbons}a, arranged by rank.  The optical gap corresponds to the discontinuity between HOMO (i= 47, -9.3 eV) and LUMO (i=48, -2.1 eV). The ionization potential is about  8.7 eV. In the ground state, all electrons are, in general, in the HOMO or below. The main change in the orbitals brought about  by defects is in the spin distribution among them, with many electrons being excited above the HOMO, and the creation of orbitals in the gap.}
\label{Fig:orbC24eV30}
\end{figure}

Figure \ref{Fig:orbC24eV30} displays, as an example, the orbital energy distribution among the 96 orbitals associated with the 24 C at. of the reference, perfectly ordered, structure of Fig. \ref{Fig:carbons} a, with single electron excitation (exceptionally) up to 30 eV.  The optical gap corresponds to the discontinuity between HOMO (i=47, -9.3 eV) and LUMO (i=48, -2.1 eV). The simulation code allows one to identify the multiplicity and the initial and final orbitals for each transition.

\begin{figure}
\resizebox{\hsize}{!}{\includegraphics{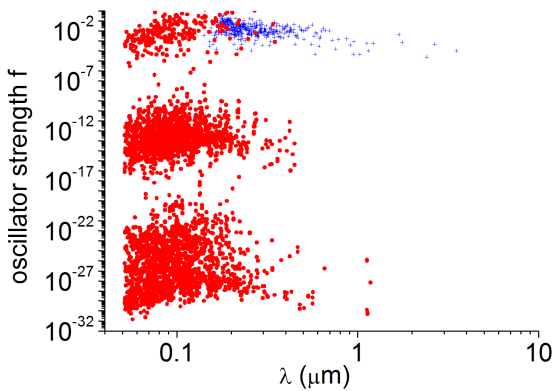}}
\caption[]{The spectral distribution of all transition strengths (logarithmic scales) for structures (a) and (b) drawn in Fig. \ref{Fig:carbons}. (a) is excited to 30 eV and (b) to 10 eV (drawn in red dots and blue crosses, respectively). Conservation of photon momentum applies to (a), not to (b). Note the large fraction of weak (forbidden) transitions in the undistorted structure (a) and their absence in the distorted one, (b). Also note the extension of the (b) spectrum into the visible and near IR.}
\label{Fig:spectraC24}
\end{figure}

Figure \ref {Fig:spectraC24} shows the spectral distributions of oscillator strengths for the reference structure Fig. \ref{Fig:carbons} a, together with that of the perturbed structure of Fig. \ref{Fig:carbons} b, singly excited up to only 10 eV.
 The momentum conservation rule applied to the former separates the 3712 transitions into 399 allowed and a majority of much weaker, forbidden, transitions; allowed transitions are very rare in the visible. By contrast, all 362 transitions in the perturbed structure are allowed, and \it they extend into the near IR. \rm The weaker excitation of the perturbed structure is responsible for the cut-off at 0.2 $\mu$m in the UV, a limitation of little consequence for our present purposes.
 
\begin{figure}
\resizebox{\hsize}{!}{\includegraphics{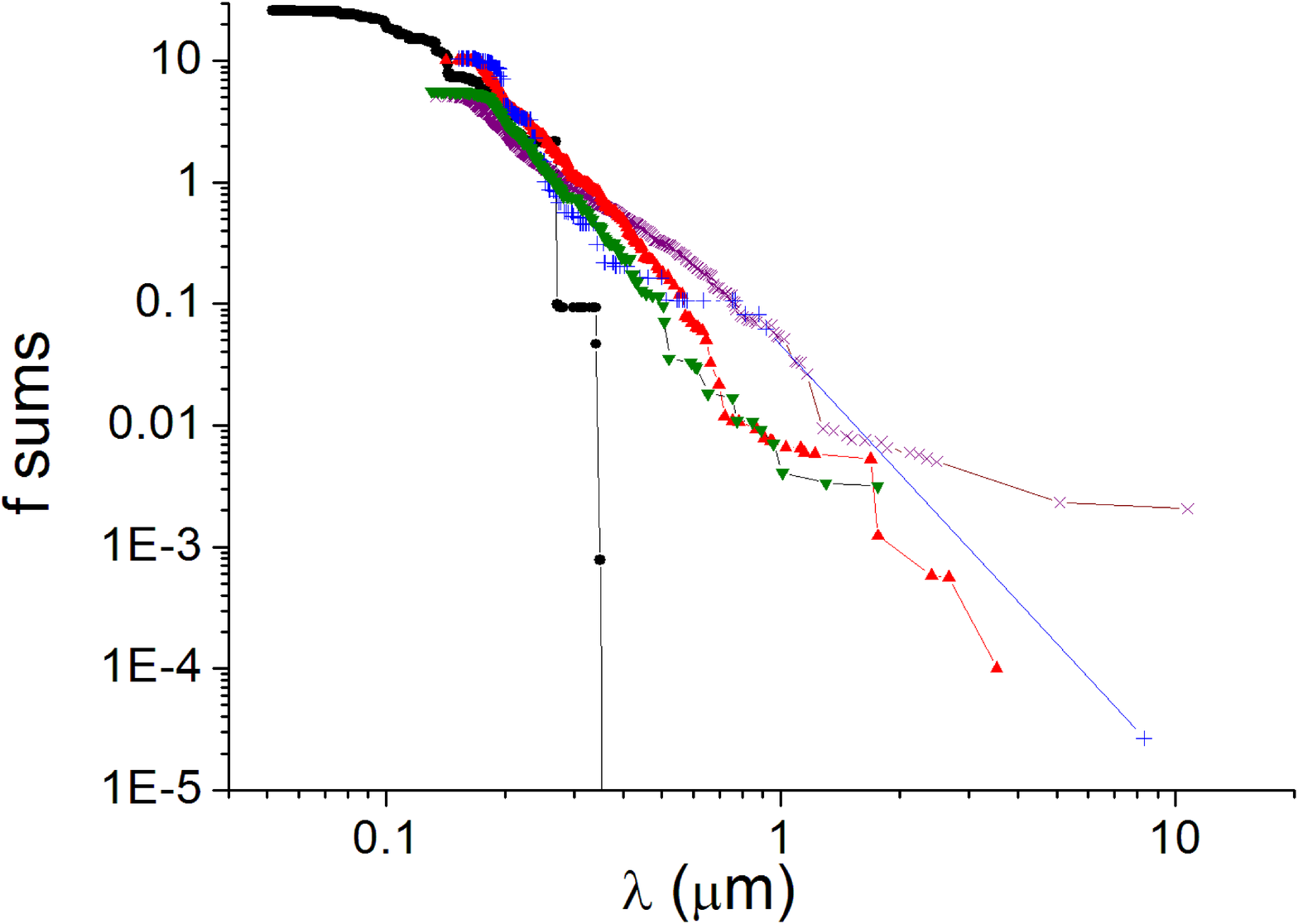}}
\caption[]{A more compact and illustrative way of presenting spectra: the running sums of oscillator strengths, f, from long to short wavelengths, for the 5 structures of Fig. \ref{Fig:carbons}: a) black dots and line, b) red triangles, c) green inverted triangles, d) blue +++, e) purple crosses. This highlights the main consequence of introducing various defects and disorder in structures: the transfer of oscillator strength from short to long wavelengths in the spectrum of transitions. Here, the only perfect structure is the coronene (a); it has a '`red'' cut-off at 0.35 $\mu$m and it extends farther into the UV because its excitation (30 eV) is stronger than that of the others (10 eV). Also, its f sum saturates at $~25$, below $\lambda=0.1 \mu$m, indicating that each C at. contributed about 1 full electron at this point, as expected for a graphitic structure. On the other hand, all the other structures are imperfect; as a consequence, the sum is more or less  depressed at short wavelengths and raised redwards as far as the near IR. This tendency is most marked with structure (e) which cumulates several defects: distortion, non-benzenic cycles, interstitial atoms and dangling bonds.}
\label{Fig:fsums}
\end{figure}

 A more compact way of representing the spectral properties of interest is to order the transitions by wavelength or energy, and sum up the corresponding strengths over the whole spectrum. This was done for the 5 structures of Fig. \ref{Fig:carbons}, and is illustrated in Fig. \ref{Fig:fsums}. This highlights the main consequence of introducing various defects and disorder in structures: the \it transfer of oscillator strength \rm from short to long wavelengths in the spectrum of transitions. Localized leaps in a curve correspond to strong individual transitions or groups of close transitions which add up to a strong one. They are more noticeable at long wavelengths because of the falling trend of the density of transitions. The number of transitions increases with grain size, all over the spectrum.
 
Here, the only perfect structure is the coronene (a); it has a '`red'' cut-off at 0.35 $\mu$m and it extends farther into the UV because its excitation (30 eV) is stronger than that of the others (10 eV). Also, its f sum saturates at $~25$, below $\lambda=0.1 \mu$m, indicating that each C at. contributed about 1 full electron up to this point, as expected for a graphitic structure. On the other hand, all the other structures are imperfect; as a consequence, the sum is more or less  depressed at short wavelengths and raised redwards as far as the near IR. This tendency is most marked with structure (e) which cumulates several defects: distortion, non-benzenic cycles, interstitial atoms and dangling bonds. As more and more such structures, oriented randomly, coalesce together, and H and C inclusions are added, the initial regularity of the graphitic structure fades away.

What is usually measured in the laboratory is the absorbance of artificially made macroscopic samples, using a spectrometer with a given spectral resolution. This is linked with f sums through Eq. 1, which applies to any single allowed transition at a given wavelength. In this, the ratio $f/\Delta\nu$ can be interpreted as the density of oscillator strength at a given wavelength. It can be deduced from the f sum by first averaging it over equal intervals corresponding to the spectrometer resolution, then differentiating this smoothed function and, in turn, smoothing the latter. Also, the volume density of absorbers, $n_{a}$, can be interpreted as the inverse of the ``volume'' $V_{a}$ occupied by our model structure. This is computed by the modeling code, taking van der Waals forces into account. Then, transforming the wavenumber interval into a wavelength interval, Eq. 1 gives the absorbance

\begin{equation}
\alpha(\lambda)=\frac{ \dot f (\lambda)\lambda^{2}}{1.2\,10^{16}\,V_{a}}\,,
\end{equation}

where $\alpha$ is in cm$^{-1}$, $\dot f$ is in $\mu$m$^{-1}$ and $V_{a}\sim10^{-21}$ cm$^{-3}$. As experimental results are often given in terms of n and k, the optical indexes of the sample, we also recall that 

\begin{equation}
\alpha=\frac{4\pi \mathrm{k}}{\lambda}= \frac{2\pi\epsilon''}{\mathrm{n}\lambda}\,
\end{equation}

\begin{figure}
\resizebox{\hsize}{!}{\includegraphics{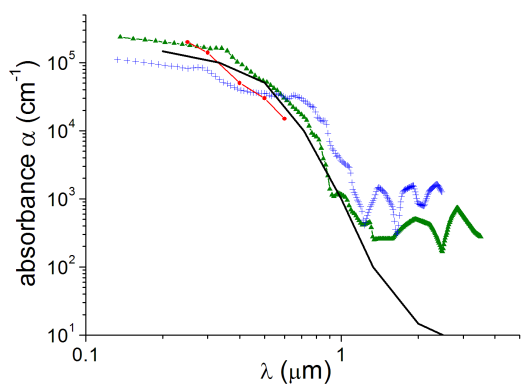}}
\caption[]{\it Green triangles:\rm\,  the computed absorbance of structure (b) in Fig. \ref{Fig:carbons}; 
\it blue +++:\rm\, the same for structure (e); \it Red line and dots:\rm\, youngest and least graphitized coal (lignite), from Ergun et al. \cite{erg}; \it black line:\rm\, a-C:H from Dischler and Brandt \cite{dis}.}
\label{Fig:Calpha}
\end{figure}

The absorbances of molecules \ref{Fig:carbons}b and e are plotted in Fig. \ref{Fig:Calpha}, together with absorbance measured on typical natural (terrestrial lignite) and artificial (a-C:H) amorphous carbons. It is apparent that, below about 1 $\mu$m, our computed models \ref{Fig:carbons} (b) and (e) fit measurements on real samples qualitatively, and quantitatively in order of magnitude. The only difference between structures (b) and (c) is the hydrogenation of the latter:  H at. insertions in the structure help distort it and fill the gap with transitions. Terminating dangling bonds with H at., as in structure (d) does not seem to add much to simple distortion. 

\it Beyond  0.5-1 $\mu$m, the absorbance curves show evidence of formation of a plateau at low levels: about 1000 cm$^{-1}$, corresponding to an imaginary index of refraction $k~$0.01. \rm This tendency is also apparent in a-C:H, but at still lower absorbance. Distortion of the structure tends to retard the set-in of the plateau and raises the level of the latter. Because the model structures are small, and the spectral density of transitions is low at long wavelengths, strong absorbance fluctuations become prominent, corresponding to fluctuations of the local density of transitions. These are reminiscent of the undulations in Fig. \ref{Fig:OpCond} d, due to heteroatomic inclusions in graphite.

\section{Silicates}

The two silicates studied here are those that were synthesized and measured in the laboratory by Scott and Duley \cite{sco}: Enstatite and Forsterite, whose chemical formula are, respectively, (MgSiO$_{2})_{3}$ and Mg$_{2}$SiO$_{4}$.

\begin{figure}
\resizebox{\hsize}{!}{\includegraphics{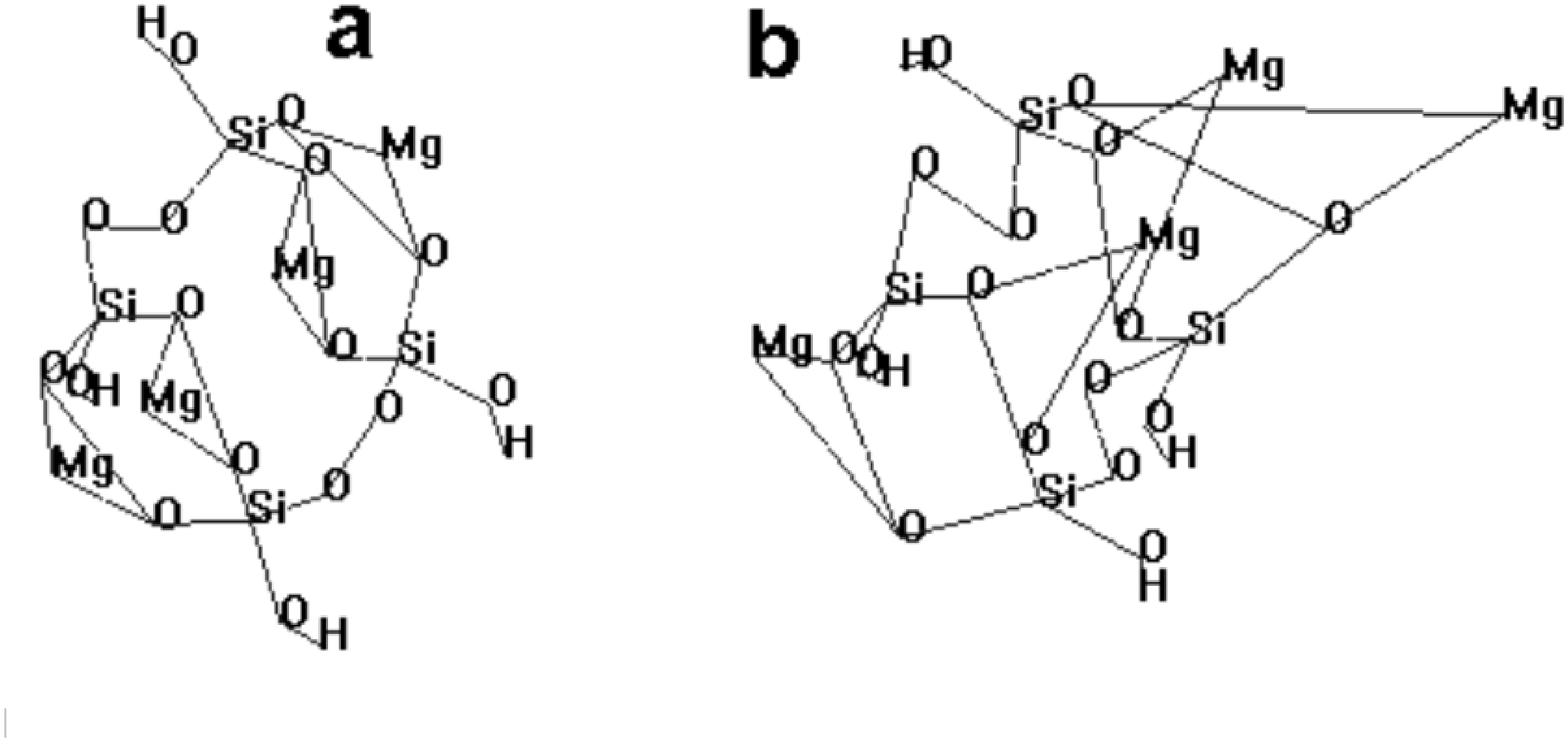}}
\caption[]{ Figure \ref{Fig:enstarticle} (a) is an optimized (unperturbed) structure (28 atoms) intended to simulate a section of an enstatite crystal according to Warren and Modell \cite{war}. Figure \ref{Fig:enstarticle} (b) is a version of (a) which has been strongly and arbitrarily distorted, to simulate an amorphous enstatite particle with 28 at.: 4 Mg, 4 Si, 16 O, 4 H.}
\label{Fig:enstarticle}
\end{figure}

Warren and Modell \cite{war} elucidated the structure of Enstatite and drew a view of it in their Fig. 4. This was taken as a model for present purposes; since the chemical modeling code can only handle a limited number of atoms, our molecules were terminated arbitrarily by severing O bonds and completing them with H atoms, as in Fig. \ref{Fig:enstarticle}. Figure \ref{Fig:enstarticle} (a) is an optimized (unperturbed) structure (28 atoms) intended to simulate a section of a crystal. Figure \ref{Fig:enstarticle} (b) is a version of (a) which has been strongly and arbitrarily distorted, and simulates an amorphous silicate particle. 

\begin{figure}
\resizebox{\hsize}{!}{\includegraphics{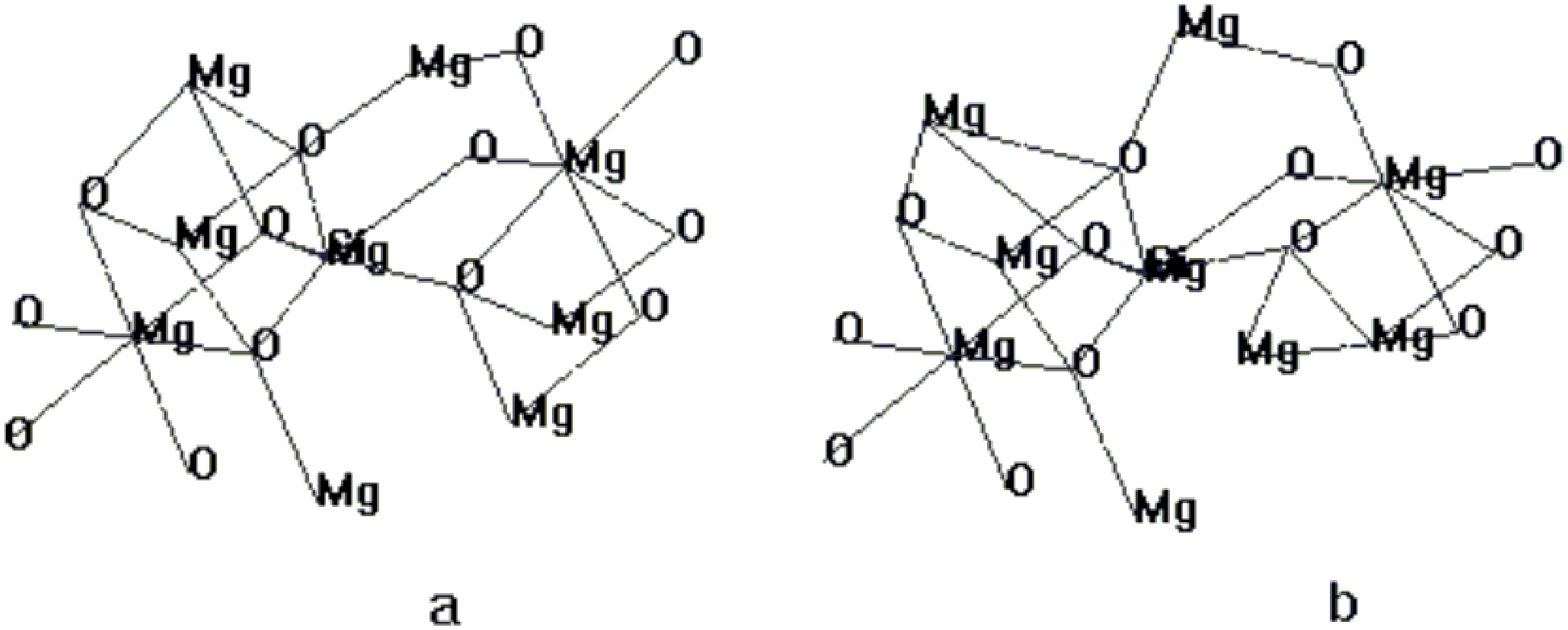}}
\caption[]{a) An elementary cell of forsterite from the MinDat  crystallographic data bank, Nb 0000276; b) the same arbitrarily perturbed by displacing individual atoms. }
\label{Fig:forstab}
\end{figure}

Similarly, Fig. \ref{Fig:forstab} represents a planar view of a) the elementary cell of crystalline Forsterite in its ground state (the unperturbed molecule), taken from the MinDat mineralogy data bank (https://www.mindat.org) 0000276; b) the same cell but distorted by displacing the atoms individually and arbitrarily.

\begin{figure}
\resizebox{\hsize}{!}{\includegraphics{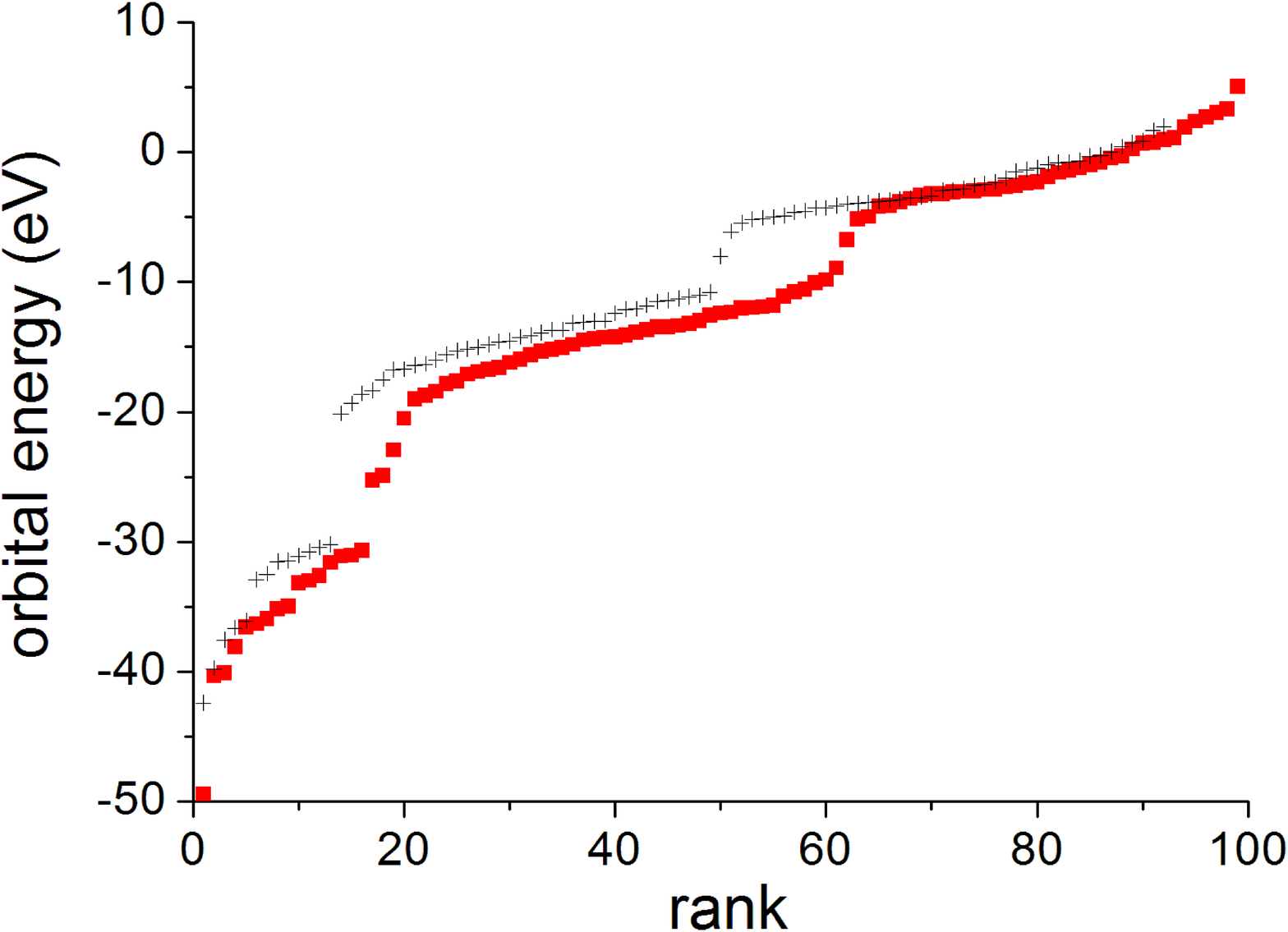}}
\caption[]{The energy distribution of orbitals for amorphous silicates; red squares: perturbed enstatite structure Fig. \ref{Fig:enstarticle} b (28 atoms); black +++: perturbed forsterite structure Fig. \ref{Fig:forstab} b (23 atoms). Comparing with Fig. \ref{Fig:orbC24eV30} for carbons, one notes the appearance, here, of orbitals in the gaps between the two valence bands and between the valence and conduction bands.}
\label{Fig:orbenstpert}
\end{figure}

Figure \ref{Fig:orbenstpert} shows energy distribution of orbitals for amorphous silicates: perturbed enstatite structure Fig. \ref{Fig:enstarticle} b (28 atoms) and perturbed forsterite structure Fig. \ref{Fig:forstab} b (14 atoms, see Sec. 4.2). Comparing with Fig. \ref{Fig:orbC24eV30} for carbons, one notes the appearance, here, of orbitals in the gaps between the two valence bands and between the valence and conduction bands

\begin{figure}
\resizebox{\hsize}{!}{\includegraphics{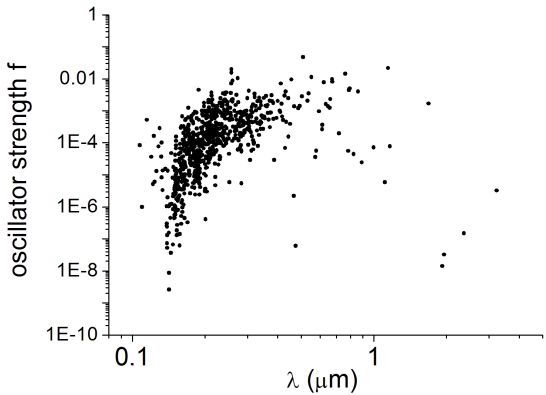}}
\caption[]{The spectral distribution of oscillator strengths of the perturbed enstatite structure, Fig. \ref{Fig:enstarticle} b. Here, again, the defects extend the spectrum into the visible and near IR, but the transitions are weaker than in the case of carbons (Fig. \ref{Fig:carbons}).}
\label{Fig:enstmuf}
\end{figure}

\begin{figure}
\resizebox{\hsize}{!}{\includegraphics{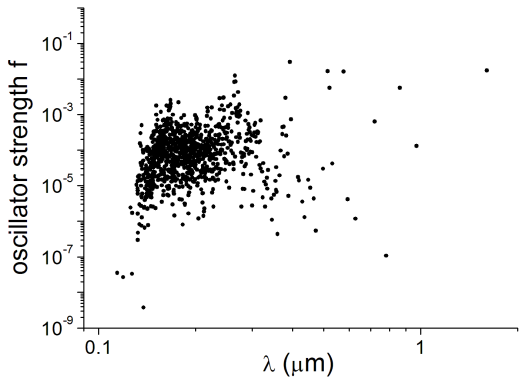}}
\caption[]{The spectral distributions of oscillator strengths of the distorted forsterite structure, Fig. \ref{Fig:forstab} b. As with carbon structures, essentially all transitions in the unperturbed cell fall below 0.4 $\mu$m, and transitions beyond only emerge when the cell is distorted.}
\label{Fig:forstmuf}
\end{figure}

\begin{figure}
\resizebox{\hsize}{!}{\includegraphics{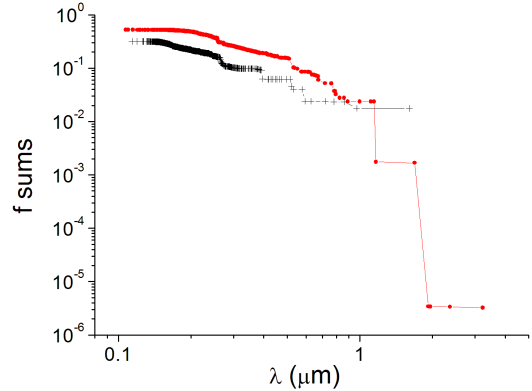}}
\caption[]{The running sums of oscillator strengths of the silicate structures Fig. \ref{Fig:enstarticle} b (red dots and line), and Fig. \ref{Fig:forstab} b (black +++ and line), to be compared with Fig. \ref{Fig:fsums} for carbons. The total sums for carbons are higher than those for silicates by more than an order of magnitude. Between $\lambda=0.1$ and 1 $\mu$m, the slopes of the running sums of carbons are much steeper than those of silicates, while the inverse is true beyond. These differences are related to the difference in conductivity between the two classes of materials.}
\label{Fig:sumssilicates}
\end{figure}

Figures \ref{Fig:enstmuf} and \ref{Fig:forstmuf} display the spectral distributions of oscillator strengths of the enstatite structure, Fig. \ref{Fig:enstarticle} b and \ref{Fig:forstab} b, respectively. Here, again, the defects extend the spectrum into the visible and near IR, but the transitions are weaker than in the case of carbons (Fig. \ref{Fig:carbons}). Figure \ref{Fig:sumssilicates} shows the running sums of oscillator strengths of structures Fig. \ref{Fig:enstarticle} b and Fig. \ref{Fig:forstab} b, to be compared with Fig. \ref{Fig:fsums} for carbons. The total sums for carbons are higher than those for silicates by more than an order of magnitude, indicating more mobility and conductivity in the former in agreement with laboratory measurements. Between $\lambda=0.1$ and 1 $\mu$m, the slopes of the running sums of carbons are much steeper than those of silicates, while the inverse is true beyond. These differences are related to the difference in conductivity between the two classes of materials.

\begin{figure}
\resizebox{\hsize}{!}{\includegraphics{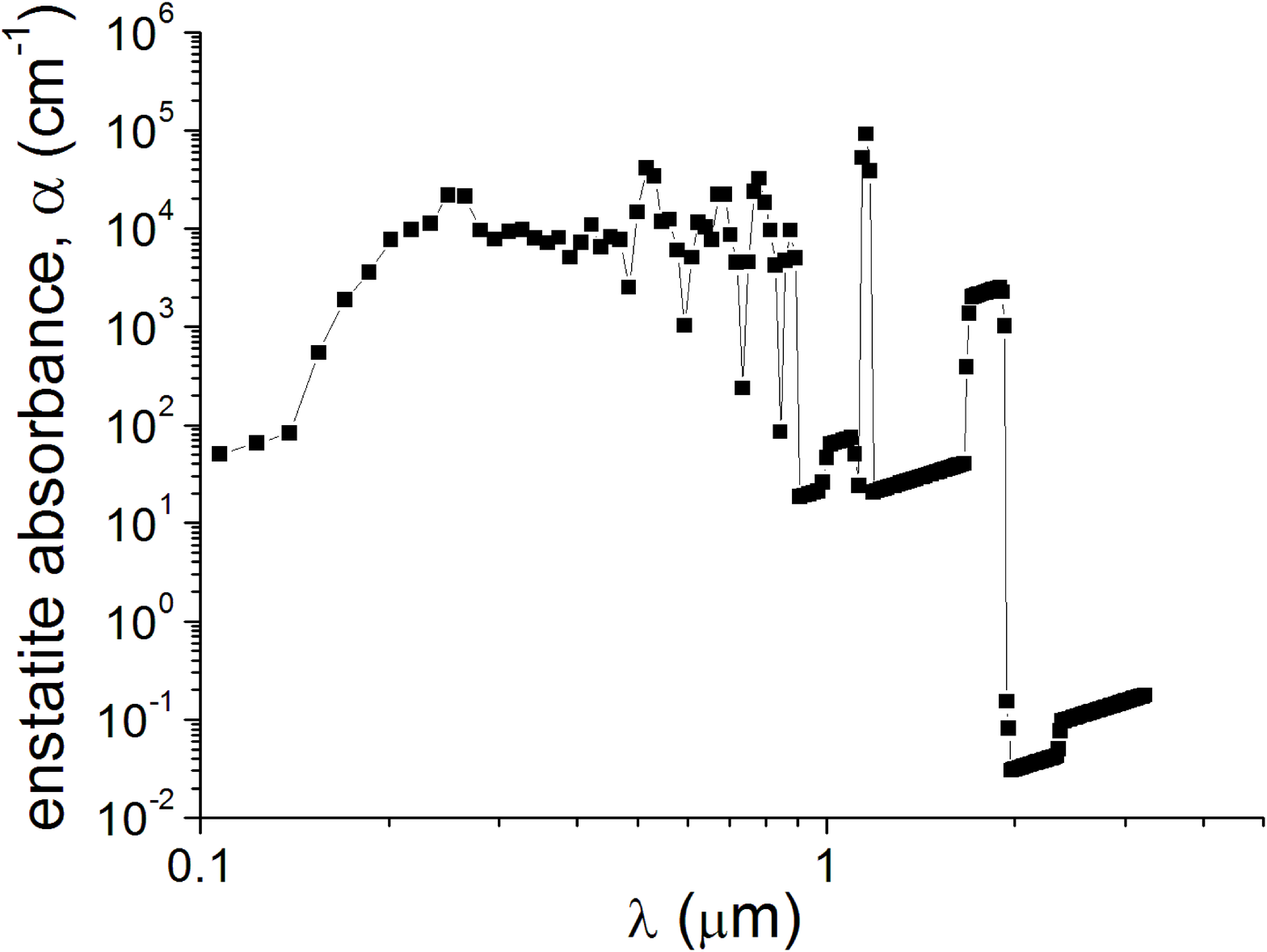}}
\caption[]{The computed absorbance, $\alpha$, of the enstatite structure Fig. \ref{Fig:enstarticle} b. Note the fall towards the UV, by contrast with the case of carbons, Fig. \ref{Fig:Calpha}. Beyond $\lambda=0.4 \,\mu$m, the scarcity of transitions favors the emergence of peaks rising above the continuum. The strongest peaks occur between 0.4 and 1 $\mu$m. They are formed by chance localized packings of strong transitions.}
\label{Fig:enstalpha}
\end{figure}

\begin{figure}
\resizebox{\hsize}{!}{\includegraphics{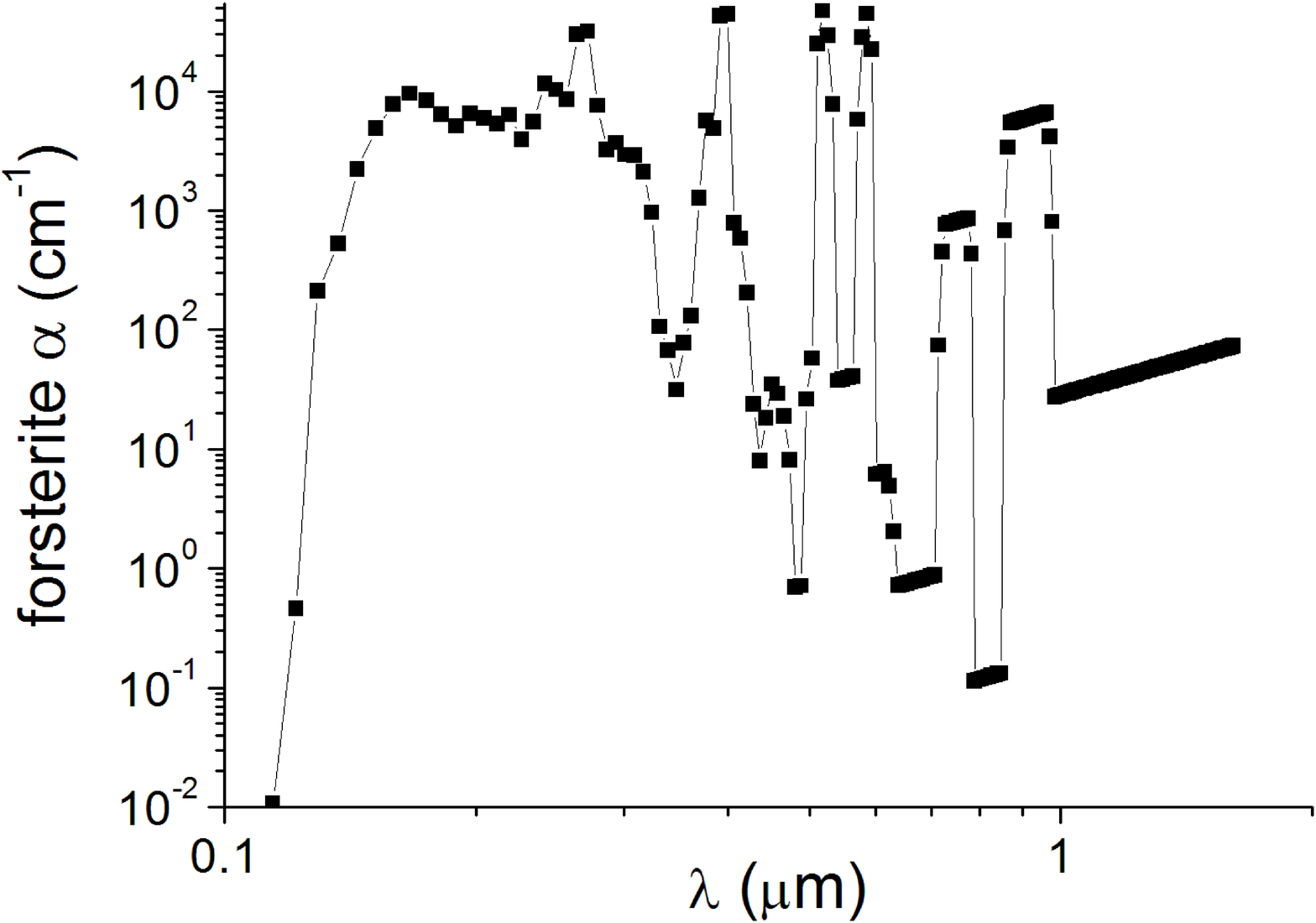}}
\caption[]{The computed absorbance, $\alpha$, of the forsterite structure Fig. \ref{Fig:forstab} b. Same comments as in Fig. \ref{Fig:enstalpha}.}
\label{Fig:forstalpha}
\end{figure}

The computed silicate absorbances, Fig. \ref{Fig:enstalpha} and \ref{Fig:forstalpha} highlight the coexistence of continuum in the UV, due to the flocking of abundant transitions, and protruding peaks in the visible, which are 
 formed by chance localized tight packings of strong transitions. As the size of the structure grows, the strength of bands relative to the underlying continuum is expected to decrease. Another remarkable fact is the fall towards the UV, by contrast with the case of carbons, Fig. \ref{Fig:Calpha}.

\begin{figure}
\resizebox{\hsize}{!}{\includegraphics{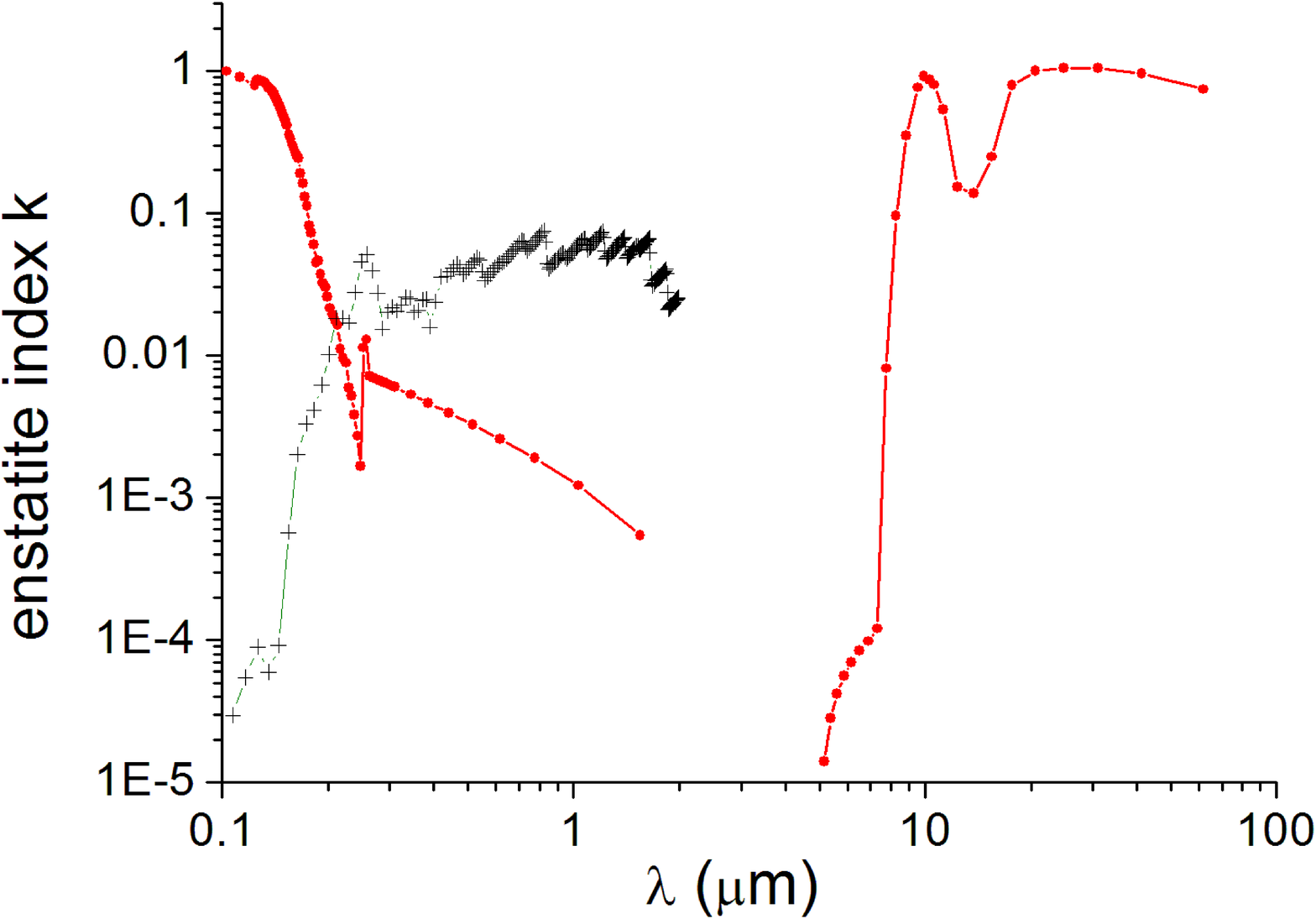}}
\caption[]{The computed imaginary index of refraction, $k$, of the enstatite structure \ref{Fig:enstarticle} b (in black crosses). The measurements of Scott and Duley \cite{sco} are superimposed (in red dots)}
\label{Fig:enstk}
\end{figure}

\begin{figure}
\resizebox{\hsize}{!}{\includegraphics{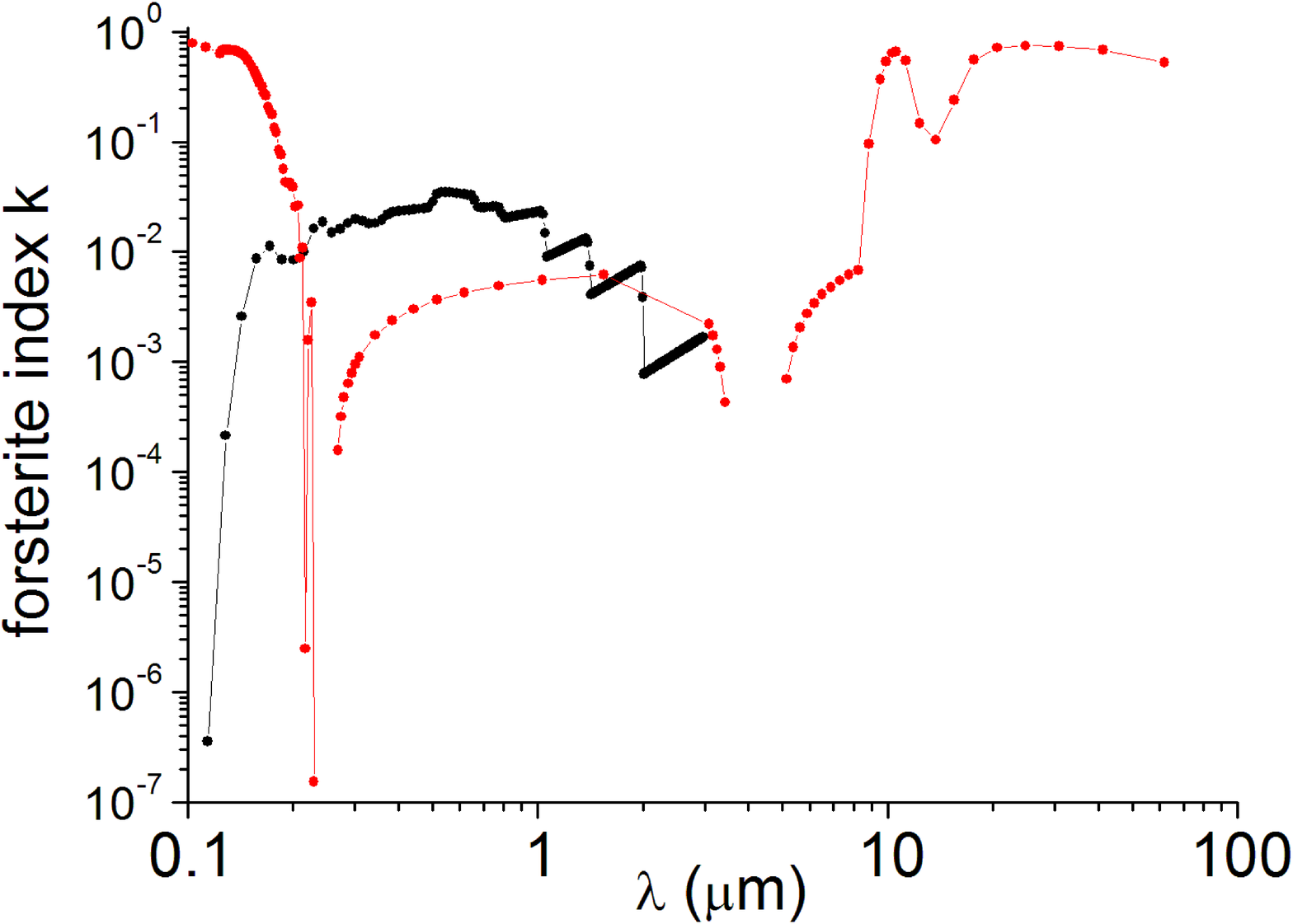}}
\caption[]{The computed imaginary index of refraction, $k$, of the forsterite structure \ref{Fig:forstab} b (in black crosses). The measurements of Scott and Duley \cite{sco} are superimposed (in red dots). For both silicates, note the steep decay of $k$towards the UV and towards the IR.}
\label{Fig:forstk}
\end{figure}

For purposes of quantitative comparison with laboratory measurements of absorbance, the f-sum from which $\alpha$ is deduced must be still more effectively smoothed than in the case of carbons, as the spectral density of bands is lower in the case of silicates. This was done to obtain the spectra of imaginary optical index $k$ displayed in Fig. \ref{Fig:enstk} and \ref{Fig:forstk}. In both cases, the continuum approximately overlays the transparency gap;  this corresponds to the finite density of states predicted by the theories of amorphous solids for the same spectral range (see Jackson et al. 1985). Our simplified model helps understand the origin of this phenomenon. The quantitative agreement with Scott and Duley's data is not satisfying for enstatite but much better for forsterite; this may be due to inadequate choice of the way the enstatite structure was distorted, but also to the fact that the values given by Scott and Duley \cite{sco} for $k$ in the gap were not directly measured, but interpolated between effectively measured ranges (in the IR and the UV) by means of the Kramers-Kronig relation.

\section{Observed and computed continuum and bands}

In as much as general trends can be inferred, and order of magnitude estimates deduced, from a limited number of relatively small carbon and silicate structures, the modeling exercises reported above suggest the following conclusions.

1) When an atomic structure is in its ground state, its binding energy is maximum, and loosely bound electrons are all packed in pairs of opposite spins in the lower, valence, orbitals. A photon can only be absorbed if it is energetic enough to cross the band gap between valence and conduction bands, whose width is typically several electron-volts. Such transitions are all in the UV. 

2) Transitions occur in the visible only when the structure is perturbed away from its ground state configuration. Perturbation includes configuration distortion, atomic inclusions in interstices, voids, dangling bonds, etc. Perturbation reduces binding energies and allows electrons into the upper, conduction, orbitals, thus making room in valence orbitals. Transitions can then occur between neighboring orbitals, differing in energy by 1 eV or less, i.e. in the visible spectral range. The larger the structure, the more orbitals are needed to describe the structure, and the closer are their energies, so the absorption spectrum extends farther into the IR. Obviously, the number of transitions increases still more rapidly towards the UV.

3)Amorphous structures may form without hydrogen: hydrogenation is only one way of favoring the dislocation of crystalline structures.

4) Our computations only provide wavelength and oscillator strength for each possible transition, not its width, which, in a solid, must be larger than the lifetime width. For particles harboring 100 atoms and more, the spectral density of transitions may be high enough to form a continuum if the instrumental resolution is not too high. This occurs more readily at shorter wavelengths, where the density of transitions is higher. 

5) The coexistence of continuum and ``bands'' in the band gap (transparency region) is more evident in silicates than in carbon structures because carbons are better conductors (see Fig. \ref{Fig:enstalpha} and \ref{Fig:forstalpha}). Both for carbons and silicates, it was shown that the computed continua agree, in order of magnitude, with laboratory measurements. 

6) If larger structures could be handled by the computer memory, the bands would be expected to drown progressively in the continuum, with only the stronger ones  emerging above, individually or, more probably, in bunches forming bands of disparate, relatively large, widths.

 \begin{figure}
\resizebox{\hsize}{!}{\includegraphics{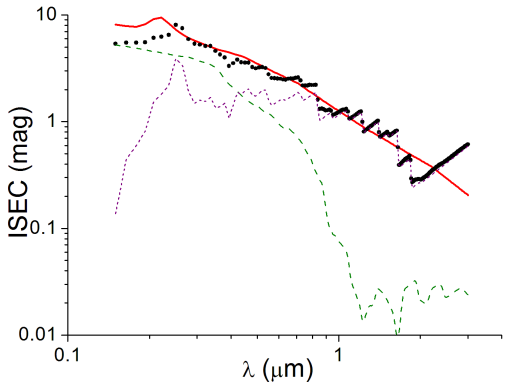}}
\caption[]{\it Red line\rm : ISEC, adapted from Mathis and Whiffen \cite{mat}, normalized to $A_{V}=3$; \it black dots \rm : tentative fit with a weighted sum of absorbances of carbon and silicate model structures studied here; \it long green dashes \rm :carbon contribution; \it short purple dashes \rm: silicate contribution. No graphite is included as its contribution is expected to be small in the visible. The kinks in the fit proceed from the discrete nature of the transitions and are expected to fade away as the grains grow. Note: the silicate contribution dominates beyond 0.4 $\mu$m.}
\label{Fig:ISEC}
\end{figure}
 
7) To check on the relevance of our computations to astronomical observations, the present computed absorbances can be confronted to the extinction observations. As an example of the latter, we take the values in table 2 of Mathis and Whiffen \cite{mat}, normalized to $E_{B-V}=1$, so $A_{V}=3$ and $N_{H}=6\,10^{21}$ cm$^{-2}$; these are plotted in red dots. The present computations are represented by the carbon structures in Fig. \ref{Fig:carbons} b and e, and the silicate structures in Fig. \ref{Fig:enstarticle} b and \ref{Fig:forstab} b, since carbons and silicates are known to be highly depleted in the ISM. Their total extinction was written in the form of a weighted sum, $\Sigma (\epsilon_{s}\alpha_{s})$, where $s$ designates the structure and $\epsilon_{s}$ its relative contribution to the sum; the latter were adjusted by cut and try to roughly fit the ISEC. No graphite absorbers were added as they are not expected to contribute much beyond 0.3 $\mu$m.

Now, for each structure, let $V_{s}$ cm$^{3}$ be its volume, $\nu_{s}$ the number of atoms of the defining element (C or Si) in the structure, $N_{s}$ cm$^{-2}$ its column density along the line of sight; let $RA_{s}$ be the abundance of its defining element  relative to H along the same line, and $fr_{s}$ the fraction of the defining element abundance that is embedded in the corresponding absorbers. Then, the contribution  of each type of structure to the total extinction may also be written

\begin{equation}
 \tau=\alpha NV= \alpha\frac{fr.RA.N_{H}.V}{\nu}
\end{equation}

where the subscript $s$ was omitted throughout. Equating this with the corresponding term of the sum $\Sigma$, one finally gets
\begin{equation}
\epsilon= \frac{fr.RA.N_{H}.V}{\nu}\,,
\end{equation}

from which $fr$ can be deduced for each structure. Here, for structures Cb, Ce, enstatite and forsterite, the volumes are, respectively, 747, 852, 850 and 676 cm$^{3}$ (i.e. an average mass density $\sim1$ g.cm$^{-3}$);  $\nu$=24, 32, 1 and 4; $RA=3.3\,10^{-4}$ for C and $3.3\,10^{-5}$ for Si; $\epsilon=1.5\,10^{-5}$, $1.5 \,10^{-5}$, $1.5\,10^{-4}$ and $2.5\,10^{-5}$. Whence, $fr$=0.24, 0.29, 1.1 and 0.61, i.e. very roughly 0.25 of all the carbon and 0.85 of all the silicon. While this is not unreasonable, it is recalled that, much earlier, Gilra \cite{gil} managed to fit the observations with a mixture of graphite, silicate and SiC, and Mathis and Whiffen \cite{mat} did so using only carbons, essentially amorphous, with a minor amount of graphite and a major volume of voids, and only mentioned the possibility of silicates without modeling these. This shows that such fits are not uneqivocal, and at best only provide rough indications. Interestingly, the present model concurs with Mathis and Whiffen's in that the contribution of graphite is small and the grains are mainly small and very porous.

\begin{figure}
\resizebox{\hsize}{!}{\includegraphics{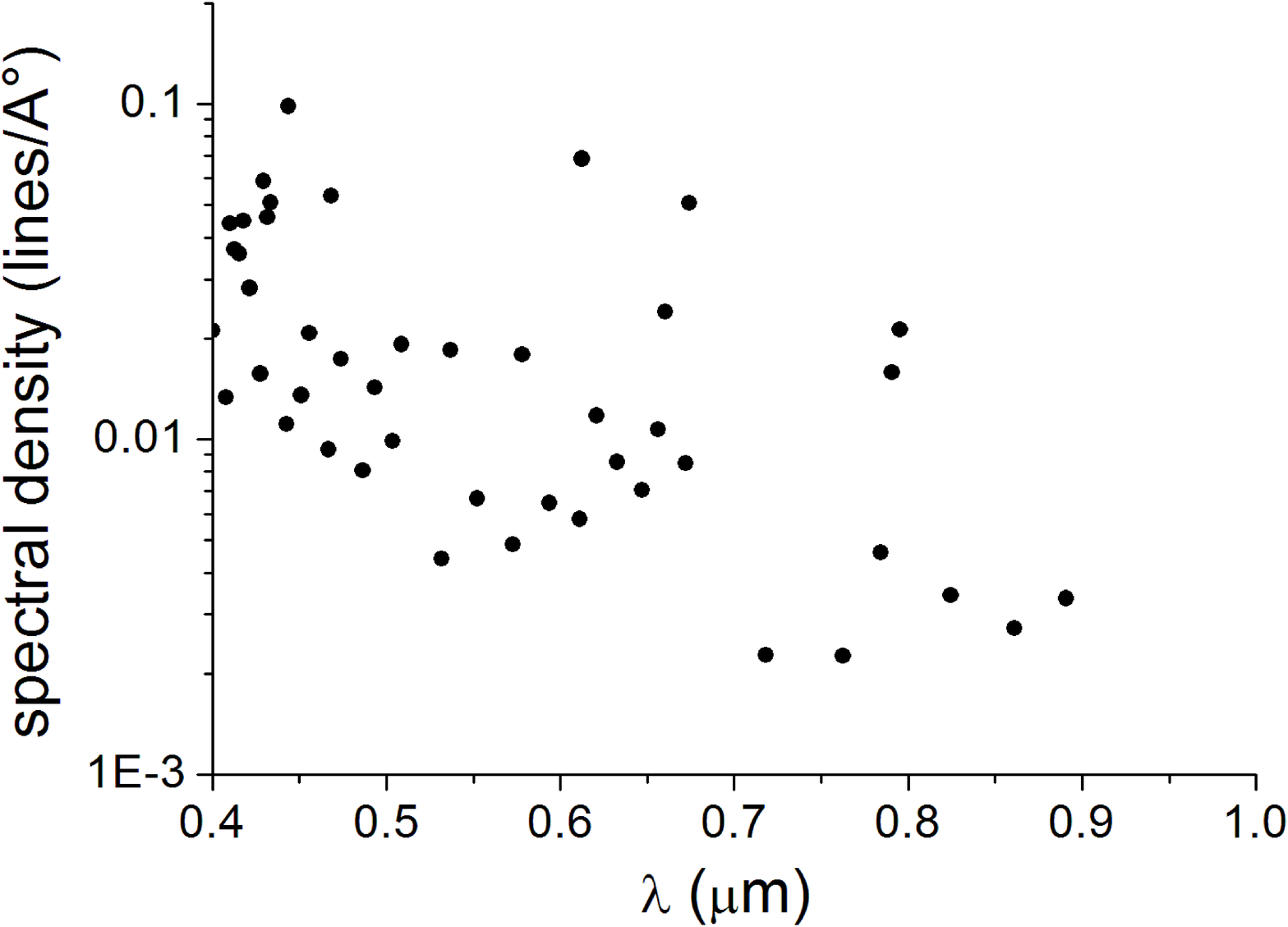}}
 \caption[]{The spectral density of the 55 visible bands of the perturbed enstatite structure of Fig. \ref{Fig:enstarticle} b. 
For each point, the abscissa is the wavelength of a computed band and the ordinate is the inverse of the interval which separates it from the next band. Density peaks of order 0.1/$\AA{\ }$ appear near 0.45, 0.65 and 0.8 $\mu$m, but most densities are of order 0.01/$\AA{\ }$. The densities increase towards the UV and decrease towards the IR.}
\label{Fig:density}
\end{figure}

8) It is widely believed that the IS grains come with a size distribution in $a^{-3.5}$ that strongly favors small sizes (see Mathis and Wallenhorst 1981). The larger grains, and the huge number of grains of all sizes along a line of sight, plausibly give rise to the continuum discussed above. However, the population of small grains cannot be ignored: it is expected that their spectra will be sparser, as the number of transitions scales like the number of atoms. Consider, for instance, the small enstatite structure of Fig. \ref{Fig:enstarticle} (28 atoms). From its computed spectrum, the spectral density (bands per \AA{\ }) can be deduced and plotted as in Fig. \ref{Fig:density}. One cannot help but compare this with the density distribution of  the 379 DIBs of Fig. 7 in Hobbs et al. \cite{hob}: this peaks at $\sim0.25/\AA{\ }$ near $\lambda=0.6\,\mu$m, with an average $\sim0.1$ over the visible. Both for DIBs and computed enstatite, the bands become very rare beyond 0.8 $\mu$m. Interestingly, the density distribution in Fig. \ref{Fig:density} is not completely random. Around the peaks, one expects the discrete lines to be observationally indistinguishable, and to add up to ``bands'', which may be stronger than the average.

 \begin{figure}
\resizebox{\hsize}{!}{\includegraphics{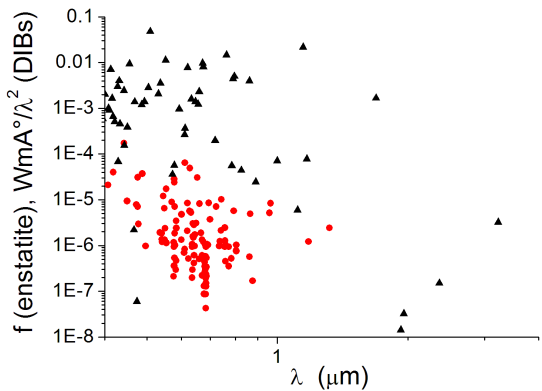}}
 \caption[]{\it Red dots\rm: Equivalent width (m$\AA{\ }$) over wavelength square ($\AA{\ }^{2}$) for 125 DIBs towards HD183143 (see Herbig 1995) ; \it black triangles\rm: the oscillator strengths, $f$,  of the perturbed enstatite  structure of Fig. \ref{Fig:enstarticle} (b) as functions of wavelength, from 0.4 to 4 $\mu$m. Both clouds of points extend over ranges of about 3 orders of magnitude but the latter is shifted upwards relative to the former by 2-3 order of magnitudes.} 
\label{Fig:parallel}
\end{figure}

9) The common origin of the continuum and lines computed above is responsible for several properties they share with  the DIBs as reported by Geballe \cite{geb11} and \cite{geb16}, and Krelowski \cite{kre} in recent reviews. The strengths of DIBs closely correlate  with $A_{V}$ over a large range (Geballe 2016, Fig. 2). DIBs are generally narrow ($\sim1\,\AA{\ }$) but many have wider, complex and asymmetric profiles. In his Fig. 4, Krelowski shows the Doppler splitting of a single DIB along one line of sight, which is interpreted as the bunching of narrow DIBs originating in different foreground clouds and having different velocities. Similarly, in the present calculations, chance flocking of stronger or weaker ETs (electronic transitions) often occur around a single frequency; at insufficient instrumental resolution, these bunches  produce the same effect. Krelowski also noted that  all DIBs should be considered as variable (his Fig. 10); this is also expected from ETs because of the variability of chemical compositions and structures of the dust along different lines of sight. 
 
Pursuing along these lines, and recalling relation 2 above, between equivalent band width and oscillator strength, Fig. \ref{Fig:parallel} displays  EW (m$\AA{\ }$) over wavelength square ($\AA{\ }^{2} $) vs. wavelength for 127 bands from HD183143 (see Herbig 1995) together with the oscillator strengths, $f$,  of the perturbed enstatite  structure of Fig. \ref{Fig:enstarticle} (b). If the model enstatite were true to the DIBs, then the two clouds of points would correspond point to point, and the ratio $f/EW$ would be the same for all pairs. Inserted in Eq. 2, this would give the column density of absorbers, $N_{a}$.

In fact, as shown in Fig. \ref{Fig:density}, the spectral density of oscillator strengths is an order of magnitude lower than that of the DIBs, which indicates that the IS DIB carriers are larger than the structures computed above. Nonetheless, both clouds of points extend roughly over equal surfaces of the plot: from the near UV to the near IR, and over about 3 orders of magnitude in ordinates; the wavelength centroid of the DIBs is in the visible and roughly coincides with the spectral range in which the computed oscillator strengths are maximum. Assuming that the size of the individual structure is increased so the number of transitions is about the same as the number of reference DIBs in the same spectral range, an estimate of the column densities of absorbers can be obtained by noting that the centroid of computed transitions is shifted upwards by 2 or 3 orders of magnitude above the DIB centroid. Inserting this result into Eq. 2, the column density of absorbers turns out to be

\begin{equation}
N_{a}\sim10^{14}-10^{15}\,\mathrm{cm}^{-2}.
\end{equation}

Much earlier, adopting 0.05 as a typical $f$ for the assumed molecular DIB carrier, Smith et al. \cite{smi} estimated this density to be approximately $3\,10^{14}$.

10) Detected DIBs show up as a small excess extinction above the continuum. In Herbig's Fig. 1, for instance, the excess is $\sim$0.2 mag over the  $\sim$3 mag of underlying visible continuum, i.e. a fraction $\sim$1/15. Assume for a moment that the DIB carriers are the relatively small grains of amorphous enstatite just described in item (9). Assume, further, as suggested by Fig. \ref{Fig:ISEC}, that the visible continuum is carried by the same amorphous material, but in larger grains, both small and large sizes belonging to the same size distribution. Since absorbance scales like the mass of the carrier, the ratio of absorbances contributed by these two sub-families of grains would be equal to the ratio of integrated masses (or volumes) over the corresponding ranges of the size distribution; this would have to be $\sim$1/15. 

In that case, if $N_{H}=6\,10^{21}$ cm$^{-2}$, as in item (7) above, then the column density of silicon \it in DIB carriers \rm  would be $\sim10^{16}$ cm$^{-2}$. Then, assuming $N_{a}=3\,10^{14}$ cm$^{-2}$ (from Eq. 6), there would be $\sim200$ atoms, on average, in each DIB carrier, i.e. 10 times more than in our silicate structures. The silicate grains carrying the continuum would still be larger by a factor 15.

Of course, the same reasoning could be applied to any other convenient amorphous material.

11) Consider an amorphous grain made of a large number of our structures, randomly coalesced; it would likely be devoid of any symmetry and could therefore hardly align with the IS magnetic field and polarize incident star light. Even if it were aligned, the lines would still be randomly polarized. This is yet another property shared with DIBs.

12) The theoretical results reported above suggest that the absorbance helps characterize a material in composition and structure because its spectral profile varies accordingly. In order to advance the identification of the IS absorbers, the work of Scott and Duley \cite{sco} needs, therefore, to be complemented by a direct and thorough comparison of extinction observations in the near UV and visible with the absorbances of several laboratory carbon and silicate samples. This requires the use of synchrotron light reflected or transmitted by very thin layers of material. With the high resolution provided by the spectrometers usually attached to such facilities, one may also hope to detect some of the strongest transitions predicted by theory near 0.4 and 0.6 $\mu$m, or between 1.1 and 1.6 $\mu$m and confront them with known DIBs (e.g. at 1.18, 1.318 and 1.5273 $\mu$m; see Geballe 2011). This applies especially to silicates, which were shown to dominate beyond 0.4 $\mu$m (Fig. \ref{Fig:ISEC}).

 \end{document}